# A contamination-free electron-transparent metallic sample preparation method for MEMS experiments with *in situ* S/TEM

Matheus A. Tunes*, Cameron Quick, Lukas Stemper, Diego S.R. Coradini, Jakob Grasserbauer, Phillip Dumitraschkewitz, Thomas M. Kremmer and Stefan Pogatscher

*Chair of Non-Ferrous Metallurgy, Montanuniversitaet Leoben, 8700 Leoben, Austria*

7 October 2020

**Abstract (200 words)**

Microelectromechanical systems (MEMS) are currently supporting ground-breaking basic research in materials science and metallurgy as they allow *in situ* experiments on materials at the nanoscale within electron-microscopes in a wide variety of different conditions such as extreme materials dynamics under ultrafast heating and quenching rates as well as in complex electro-chemical environments. Electron-transparent sample preparation for MEMS e-chips remains a challenge for this technology as the existing methodologies can introduce contaminants, thus disrupting the experiments and the analysis of results. Herein we introduce a methodology for simple and fast electron-transparent sample preparation for MEMS e-chips without significant contamination. The quality of the samples as well as their performance during a MEMS e-chip experiment *in situ* within an electron-microscope are evaluated during a heat treatment of a crossover AlMgZn(Cu) alloy.

\* *Corresponding author:* m.a.tunes@physics.org

## 1. Introduction

The use of microelectromechanical systems (MEMS) within scanning/transmission electron microscopes (S/TEM) is at the forefront of experimental science, particularly in the fields of nanotechnology and materials science. As a rapidly evolving and emerging technology, MEMS experiments with *in situ* TEM can provide countless opportunities for investigation of the real-time response of materials in corrosive and gaseous environments (Noh et al., 2011; Zhong et al., 2016), under extreme dynamic changes when subjected to ultrafast heating and cooling rates of up to $10^6$ K·s$^{-1}$ (Damiano et al., 2008; Novák et al., 2016), under mechanical loading (Hattar et al., 2004; Tochigi et al., 2019) or when subjected to complex photocatalytic environments (Dillon & Liu, 2012; Sun, 2020). These experiments are now in fact contributing to the design of new materials at the nanoscale as well as supporting the progress of basic research in science by allowing complex physicochemical (Li et al., 2020) and/or elastoplastic (Imrich et al., 2015) mechanisms to be fundamentally investigated at the nanoscale.

A major challenge when carrying out MEMS experiments with *in situ* TEM lies in the sample preparation methodology chosen for producing good-quality, electron-transparent lamellae and their subsequent transfer to the MEMS e-chips. Up to now, such sample preparation methodology is highly dependent on the application of dual-beam scanning electron microscopes (SEM) with focused ion beam (FIB) capabilities (Tuck et al., 2004; Novák et al., 2016; Vijayan et al., 2017).

The unquestionable reliability and efficiency are evident characteristics of FIB-based methods for producing electron-transparent samples from metallic substrates (Giannuzzi & Stevie, 1999), but several degradation mechanisms are reported to occur during the stages sample preparation within SEM-FIBs (Mayer et al., 2007). These may impact the final quality of a specimen and possibly affect the reliability of the results generated during MEMS experiments within a S/TEM. Many metallurgical samples, such as those made from Al-based alloys, may strongly interact with either Ga ions (Ernst et al., 2017) or Pt/C layers often used as a top-protective coatings (Mayer et al., 2007; Bender et al., 2015) resulting in contamination and subsequent formation of undesirable artefacts. The advent of plasma-based FIBs (using Xe ions) is reported to mitigate some deleterious effects found in Ga-based FIBs (Ernst et al., 2017), but the use of Xe ions can impact electron-transparent metallic lamellae in different ways including radiation-induced damage or even the formation of nanometre-sized Xe bubbles (Estivill et al., 2016).

Given the facts, the optimal scenario would be an electron-transparent sample preparation methodology free of both contaminants (Ga, Pt or Xe) and radiation-induced damage effects. We report in this paper an alternative methodology for producing good-quality and contaminant-free electron-transparent metallic specimens for MEMS experiments with *in situ* S/TEM. Following a detailed description of the MEMS electron-transparent sample preparation methodology, a heat treatment experiment *in situ* within a S/TEM is presented using a MEMS chip and the quality of the produced specimen before and after the experiment is evaluated using conventional and analytical electron-microscopy techniques.

## 2. Materials and Methods

### a. *Provenance of the metallic samples*

The MEMS sample preparation methodology reported in this research works for a wide variety of metallic samples, but for the experiments reported in this present paper a novel crossover $AlMg_{4.7}Zn_{3.6}Cu_{0.6}$ alloy (in wt.%) was used. Throughout the text, these samples will be referred to as "crossover AlMgZn(Cu) alloy". The alloy was pre-aged for 3h at 373 K and subjected to a minor deformation level of 2% (for details on the synthesis, processing and materials properties we refer to (Stemper et al., 2020)).

**b.** *Jet Electropolishing (JEP)*

Electron-transparent specimens of the crossover AlMgZn(Cu) alloy were prepared using the technique of jet electropolishing. The samples were polished and ground to 100 µm of thickness and punched out to 3 mm disks. For the JEP procedure, an electrolyte solution composed of 25% nitric acid and 75% methanol (in vol.%) was used at a temperature range of 253–257 K with the electrode potential set to 12 V. During JEP, the specimen current slightly oscillated around 90 mA. After JEP, the samples were washed in three different pure methanol baths and left to dry in air.

**c.** *Scanning/Transmission Electron Microscopy (S/TEM)*

Electron-microscopy was carried out using a Thermo Fisher Scientific™ Talos F200X scanning/transmission electron microscope. The microscope operates a X-FEG filament (a refinement of the Schottky thermally assisted field emission gun) at 200 kV and features the Super-X energy dispersive X-ray (EDX) spectroscopy technology. For the investigations reported in this paper, the following imaging modes were used: bright-field TEM (BF-TEM), selected-area electron diffraction (SAED), low-angle annular dark-field (LAADF) and bright-field STEM (BF-STEM).

**d.** *MEMS experiments with in situ S/TEM*

*In situ* S/TEM heat treatment experiments were performed using a Protochips FUSION 200 MEMS chip-based holder with double-tilt capability. For the heat treatment experiments reported in this work, the crossover AlMgZn(Cu) alloy was subjected to a heating ramp of +60 K·min$^{-1}$ up to 458 K where the samples were held for 1200 s. Then, a cooling ramp of –60 K·min$^{-1}$ was applied down to a temperature of 298 K. This heat treatment specification is denoted in the metallurgical literature as a paint bake (Stemper et al., 2020). For the MEMS experiments reported in this work, e-chips without coating on the SiN membrane (a hollow region with 9 holes where the electron-transparent piece was placed to be analysed within the S/TEM) were used.

## 3. Results and Discussion

### a. MEMS sample preparation methodology

The MEMS sample preparation methodology investigated in this work is described in the set of optical micrographs in Figures 1(a-i). The entire process can be performed using a simple stereo microscope with magnification in the order of 100-200x.

The methodology consists of pre-selecting an electropolished 3 mm disk that was pre-confirmed to have electron-transparent areas around its central hole as shown in Figure 1(a). This central hole has the regions-of-interest (ROI) for the MEMS experiments with *in situ* S/TEM. The 3 mm disk is placed onto a glass slide then, with a sharp laboratory scalpel, a series of cuts are performed on the 3 mm disk in order to isolate parts of the ROI as shown in the set of optical micrographs in Figures 1(b-e). The samples prepared in this work used a polished sapphire slide as a cutting surface, however conventional laboratory glass slides are adequate. For optimal results when making the cuts, the authors advise using a curved scalpel, and aligning the intended cut while the scalpel tip is in contact with the cutting surface. Firmly lowering the scalpel handle from this position provides the best control, and allows the precision necessary to isolate the ROI. With ROI cut into smaller pieces of around 50-100 μm as denoted, one piece is selected to be transferred onto the MEMS chip as shown in the inset of the optical micrograph in Figure 1(f).

With the electron-transparent piece (≈50 μm) cut and selected from the ROI, the MEMS chip is placed into the field-of-view: the membrane of the MEMS chip can be seen in the stereo microscope as shown in Figure 1(g) and this is the target area for the electron-transparent piece.

The transfer procedure is performed with the use of an animal hair (taken from a regular high-quality paint brush commonly found in stationery shops). The hair is then statically charged by friction and its tapered point is used to catch the electron-transparent piece and deposit it onto the membrane of the MEMS chip as shown in the micrographs of Figure 1(h-i). For slight repositioning of the sample upon the membrane, the hair tool can be washed in isopropanol. This serves to remove any residual static charge and allows manipulation of the sample without it attaching to the hair. The electron-transparent piece sticks firmly onto the MEMS e-chips and empirical experience (acquired by repeating this process several times) shows that smaller samples (on the order of the membrane dimensions of the MEMS chip, *i.e.* 50x50 µm) do not fall from the MEMS chip during sample loading into the electron-microscope, despite the holder is turning upside-down during the loading steps. With practice, the entire process takes only about 15 min.

A final note must be made regarding the sample preparation methodology above described. The final quality of the specimen will be dependent on the initial quality of the electropolished 3 mm disks as well as on overall cutting procedure. Although brittle thin samples can be easily cut with a scalpel, the crossover AlMgZn(Cu) alloy used in this work is highly ductile (Stemper et al., 2020) and cutting was also easily performed. Therefore, it is expected that the methodology works for electropolished metallic samples either brittle or ductile. Ceramic materials were not tested in this work, but given that 3 mm disks can be punched ultrasonically and subsequent mechanically thinned, dimpled and ion-polished (using a precise ion polishing system or PIPS) to electron-transparency, the proposed methodology may also be applicable for this class of materials.

### b. *Characterisation with S/TEM*

The electron-transparent piece attached to the MEMS e-chip as shown in Figure 1(i) was loaded into the electron-microscope. Figure 2(a) shows a BF-TEM micrograph taken at low-magnification indicating that the sample was stationary on the MEMS chip during both loading procedures into the holder and into the electron-microscope. Figure 2(a) also indicates that the sample is covering four holes of the membrane of the MEMS e-chip. A high-magnification BF-TEM micrograph was taken from the hole indicated with a blue square in Figure 2(a): as shown in Figure 2(b), the sample is high-quality and electron-transparent, thus demonstrating the viability of the proposed sample preparation methodology.

It is worth emphasising that when using the SEM-FIB for producing samples for MEMS e-chips, Pt is often deposited in specific areas of the sample in order to weld it onto the e-chip membrane (Zhong et al., 2016). This step may introduce a significant yield of Pt contamination onto the surfaces of the electron-transparent specimen which may compromise the MEMS experiments. The same is expected to occur upon interaction of either Ga or Xe ion beams with the metallic specimens during the stages of trenching, milling, cutting and polishing within the SEM-FIB. Conversely, in the several steps of the sample preparation methodology reported in this work neither Pt, Ga nor Xe are used, therefore the sample will be contaminant-free when compared to those made within the SEM-FIB.

c. *Paint bake of a crossover AlMgZn(Cu) alloy in situ within a S/TEM*

In order to evaluate both quality and performance of the electron-transparent sample during a MEMS experiment, a paint bake treatment was performed *in situ* in the STEM which consisted of heating the crossover AlMgZn(Cu) alloy up to 458 K for 1200 s. The BF-STEM and LAADF micrographs in Figures 3(a) and 3(f) show the microstructure of the alloy prior to paint bake which is composed of nanometre-sized Guinier-Preston zones (or simply GP Zones) and a high density of dislocations given that the alloy was 2% deformed. Note that this is similar microstructure to that found in a conventionally produced AlMgZn(Cu) alloy (Stemper et al., 2020).

Upon paint bake, the microstructural evolution of the crossover AlMgZn(Cu) alloy was monitored real-time using the BF-STEM and LAADF detectors as shown in the micrographs in Figures 3(b-e) and Figures 3(g-j), respectively. A reorganization of the initial dislocation structure was observed to take place including complete annihilation of pre-existing dislocations. During the experiment, the sample experienced minimal drift, only on the order of few nanometres. This can be better visualised with a video record showing the complete microstructural evolution during the experiment and which is provided as supplementary information.

The BF-STEM micrographs and SAED patterns shown in Figure 4(a-b) and 4(c-d), respectively, exhibit the microstructure of the crossover AlMgZn(Cu) alloy before and after the paint bake treatment. The GP zones are still present in the microstructure although some of the pre-existing dislocations in Figure 4(a) did undergo annihilation as mentioned before. It is worth emphasising that the both SAED patterns before and after the experiment do not shown any Debye-Scherrer ring commonly associated with polycrystalline nanometre-sized artefacts introduced by Pt and Ga contamination when using samples produced via the SEM-FIB technique. As this is an experiment performed with an electron-transparent lamellae at the nanoscale, the results may not correspond directly to those reported by Stemper where heat treatment was performed in bulk specimens (Stemper et al., 2020).

### d. *Post-experiment impurity analysis*

STEM-EDX was used to analyse the elements present in the sample after the *in situ* STEM paint bake treatment. Figure 5 shows a long-exposure (2 h) STEM-EDX raw spectrum collected from the whole area corresponding to the micrograph in Figure 4(b). The intensity axis (y-axis) of the STEM-EDX plot in Figure 5 was set to logarithmic scale in order to better evaluate the presence of minor elemental peaks and a locally estimated scatterplot smoothing (LOESS) fit was used to better identify the peaks' positions in the energy axis (x-axis). In order to define the relevance of a peak with respect to the background noise, the relative intensity of each identified peak was calculated against the strongest signal peak, *e.g.* the Al Kα peak index 7 located at 1.487 keV. Peaks with less than 0.01% of relative intensity were not considered.

Using this methodology, 20 peaks were identified in the STEM-EDX raw spectrum. The peaks' indexes, as well as their energy and relative intensity, were extracted from the plot. The results are shown in Table 1 and the accuracy of the STEM-EDX detector is noted. Most of the peaks are confirmed to have an energy match (with their expected position using reference values in the software Velox) in the third decimal place. Given such accuracy, the absence of contaminants such as Pt (Lα = 9.442 and Mα = 2.050 keV) and Ga (Kα = 9.251 and Lα = 1.098 keV) is remarkably noted. Minor impurities were identified as coming from the alloy production: Ca and Zr with extremely low relative intensity, 0.13 and 0.01% respectively. The presence of O is expected as Al self-passivates. Due to overlap between multiple different elements, the peaks corresponding to labels 3, 9, 10, 12, 17 and 18 were not properly identified, although their relative intensity is below 1%; therefore, we assume these impurities are from the inner microscope electronics, holder, e-chip or background noise.

The only contaminant observed with relative intensity of 1.6% was the element C (peak label 1). It is well known that EDX precludes the identification of the element C (Isabell et al., 1999), but the presence of this small peak can indeed be attributed to carbonaceous contamination either in the surface of the crossover AlMgZn(Cu) sample or in the MEMS e-chip. However, this C contamination can be mitigated with the use of plasma cleaning which was not applied in this work. Regardless of the minor C contamination and the presence of small impurities, the experiment and its outcomes were not in any way affected as confirmed by the detailed post paint bake electron-microscopy analysis. No artefacts were observed to nucleate and grow on the alloy microstructure as a result of the experiment.

## 4. Summary and conclusions

An alternative method for producing good-quality and contaminant-free electron-transparent samples for MEMS experiments *in situ* within a S/TEM was introduced in this paper. The method consisted of using electropolished 3 mm disks from metallic samples with an electron-transparent hole in the centre. The disk is then subjected to a set of precise cuts in order to separate the electron-transparent region into smaller pieces of around ≈50 µm. The electron-transparent piece can be transferred to pristine MEMS e-chips with a high-quality animal hair used as a micrometre-sized manipulation tool.

The introduced methodology is faster than SEM-FIB and it allows the sample preparation of multiple samples from only one 3 mm electropolished disk. A paint bake experiment of a crossover AlMgZn(Cu) alloy was performed in order to attest the quality of the sample and its stability during an *in situ* STEM experiment. Yield of minor impurities were observed to come from the Al alloy itself rather than the sample preparation method. The only minor extrinsic contamination observed was C which is commonly identified when using the EDX technique for elemental estimation. None of the minor impurities nor C contamination were observed to affect the results of the paint bake experiment as no artefacts were observed to form and evolve in the sample.

This sample preparation methodology works well for both ductile and brittle metallic 3 mm electropolished disks, but it has not been yet tested for 3 mm dimpled and ion-polished ceramic discs.

## 5. Acknowledgements

The European Research Council (ERC) excellent science grant "TRANSDESIGN" provided funding for the research reported in this work through the Horizon 2020 program under contract 757961. The electron-microscopy facility used in this work received funding from the Austrian Research Promotion Agency (FFG) project known as "3DnanoAnalytics" under contract number FFG-No. 858040.

## 7. Tables

**Table 1:** STEM-EDX impurity analysis of the paint baked crossover Al alloy.

| Peak Index | Measured [keV] | Expected [keV] | Relative‡ [%] | Identified Element |
|---|---|---|---|---|
| **1†** | 0.273 | 0.280 | 1.60 | C Kα |
| **2** | 0.521 | 0.524 | 3.20 | O Kα |
| **3†** | 0.675 | - | 0.61 | Multi. Elements |
| **4** | 0.933 | 0.929 | 2.89 | Cu Lα |
| **5** | 1.018 | 1.012 | 3.34 | Zn Lα |
| **6** | 1.253 | 1.254 | 6.34 | Mg Kα |
| **7** | 1.487 | 1.487 | 100 | Al Kα |
| **8** | 1.742 | 1.742 | 2.31 | Si Kα |
| **9†** | 2.623 | - | 0.21 | Multi. Elements |
| **10†** | 2.971 | - | 0.18 | Multi. Elements |
| **11†** | 3.699 | 3.691 | 0.13 | Ca Kα |
| **12†** | 5.910 | - | 0.08 | Multi. Elements |
| **13** | 8.055 | 8.048 | 0.09 | Cu Kα |
| **14** | 8.644 | 8.639 | 0.09 | Zn Kα |
| **15** | 8.921 | 8.907 | 0.14 | Cu Kβ |
| **16** | 9.587 | 9.574 | 0.14 | Zn Kβ |
| **17†** | 10.558 | - | 0.04 | Multi. Elements |
| **18†** | 12.643 | - | 0.03 | Multi. Elements |
| **19†** | 15.764 | 15.775 | 0.04 | Zr Kα |
| **20†** | 17.671 | 17.667 | 0.01 | Zr Kβ |

†Note 1: all the identified impurities have low relative intensity (*i.e.* the signal from impurities are comparable to noise).

‡Note 2: the relative intensity was calculated with respect to the most intense peak in the spectrum (the Al Kα peak index 7).

## 8. Figure Legends

**Figure 1:** Step-by-step description of the sample preparation methodology proposed in this work. The optical micrograph in (a) shows the 3 mm disk of the electropolished crossover AlMgZn(Cu) alloy with a central hole where the electron-transparent regions are located. Steps (c-e) show the subsequent cutting made with a sharp laboratory scalpel: note the cuts are made to separate the regions-of-interest around the central hole from the whole 3 mm disk. The optical micrograph in (f) shows three pieces of an electron-transparent area that have been cut: a mid-sized piece of size around 50 µm as indicated by the yellow arrow in the inset in (f) was selected for transfer onto the MEMS chip. The optical micrograph in (g) shows the MEMS chip without the sample in its membrane (with holes). The sample is then transferred (h) to the MEMS chip by using a piece of brush-bristle with intrinsic static after frictional static charging. The sample is positioned on the MEMS chip membrane as shown in (i). The leftover pieces (from the inset in (f)) can still be used to produce additional samples in different MEMS e-chips.

**Figure 2:** BF-TEM micrographs after the sample preparation procedure showing (a) an electron-transparent sample of the crossover AlMgZn(Cu) alloy attached to the MEMS chip and (b) the sample lying over a hole (indicated by the blue square in (a)) – on the MEMS chip.

**Figure 3:** Heat treatment of the crossover AlMgZn(Cu) alloy *in situ* within a S/TEM using the MEMS chip. The set of micrographs (a-e) and (f-j) are showing the microstructural evolution of the crossover AlMgZn(Cu) alloy as a function of time with the BF-STEM and LAADF detectors, respectively. Micrographs (a) and (f) were taken prior paint bake whilst micrographs (b,g), (c,h), (d,i) and (e,j) were taken at 200, 400, 800 and 1200 s, respectively.

**Figure 4:** BF-STEM and SAED patterns of the crossover AlMgZn(Cu) alloy oriented along [112] zone-axis (a-c) before and (b-d) after paint bake treatment. The additional spots in (c) and (d) are due to dispersoid phases not in the field of view.

**Figure 5:** STEM-EDX raw spectrum acquired from the whole area covered in micrograph 4(b). Note: the y-scale was set to logarithm in order to maximize visualisation smaller peaks.

## 9. Figures

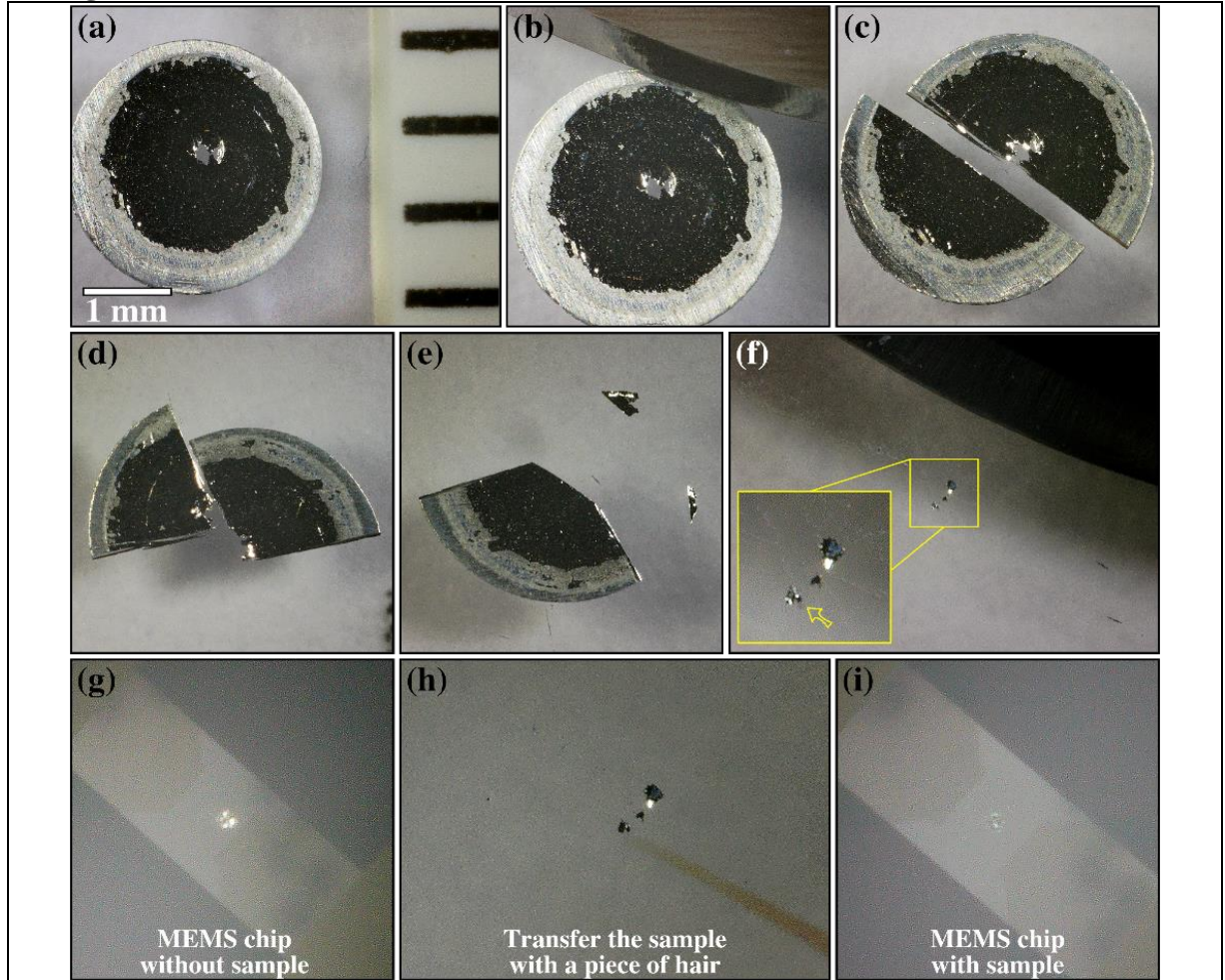

**Figure 1:** Step-by-step description of the sample preparation methodology proposed in this work. The optical micrograph in (a) shows the 3 mm disk of the electropolished crossover AlMgZn(Cu) alloy with a central hole where the electron-transparent regions are located. Steps (c-e) show the subsequent cutting made with a sharp laboratory scalpel: note the cuts are made to separate the regions-of-interest around the central hole from the whole 3 mm disk. The optical micrograph in (f) shows three pieces of an electron-transparent area that have been cut: a mid-sized piece of size around 50 µm as indicated by the yellow arrow in the inset in (f) was selected for transfer onto the MEMS chip. The optical micrograph in (g) shows the MEMS chip without the sample in its membrane (with holes). The sample is then transferred (h) to the MEMS chip by using a piece of brush-bristle with intrinsic static after frictional static charging. The sample is positioned on the MEMS chip membrane as shown in (i). The leftover pieces (from the inset in (f)) can still be used to produce additional samples in different MEMS e-chips.

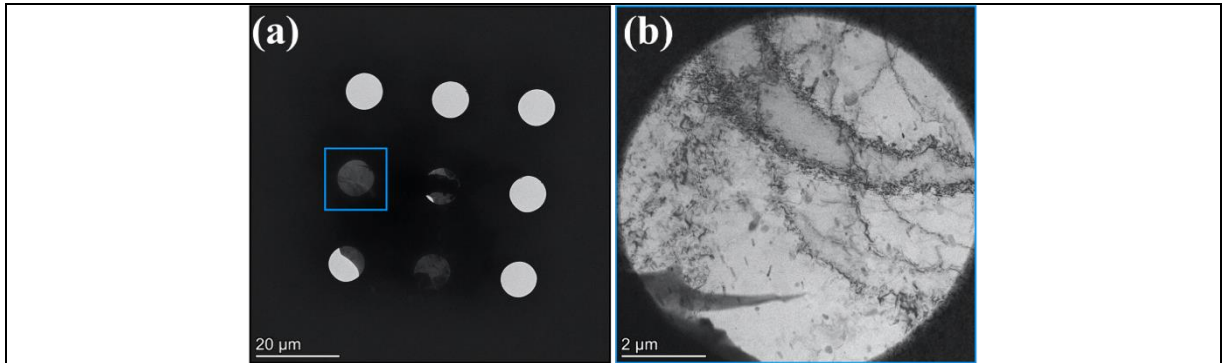

**Figure 2:** BF-TEM micrographs after the sample preparation procedure showing (a) an electron-transparent sample of the crossover AlMgZn(Cu) alloy attached to the MEMS chip and (b) the sample lying over a hole (indicated by the blue square in (a)) – on the MEMS chip.

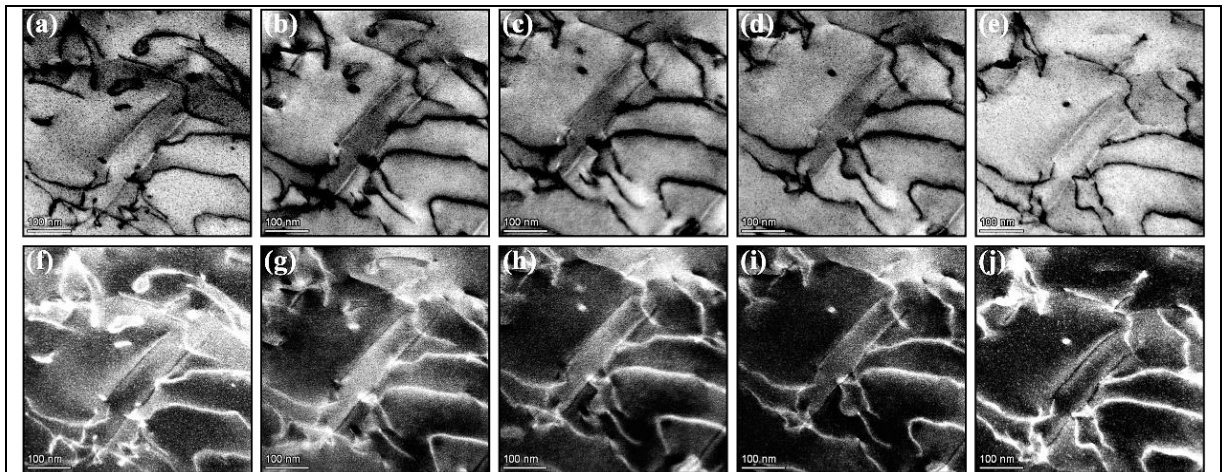

**Figure 3:** Heat treatment of the crossover AlMgZn(Cu) alloy *in situ* within a S/TEM using the MEMS chip. The set of micrographs (a-e) and (f-j) are showing the microstructural evolution of the crossover AlMgZn(Cu) alloy as a function of time with the BF-STEM and LAADF detectors, respectively. Micrographs (a) and (f) were taken prior paint bake whilst micrographs (b,g), (c,h), (d,i) and (e,j) were taken at 200, 400, 800 and 1200 s, respectively.

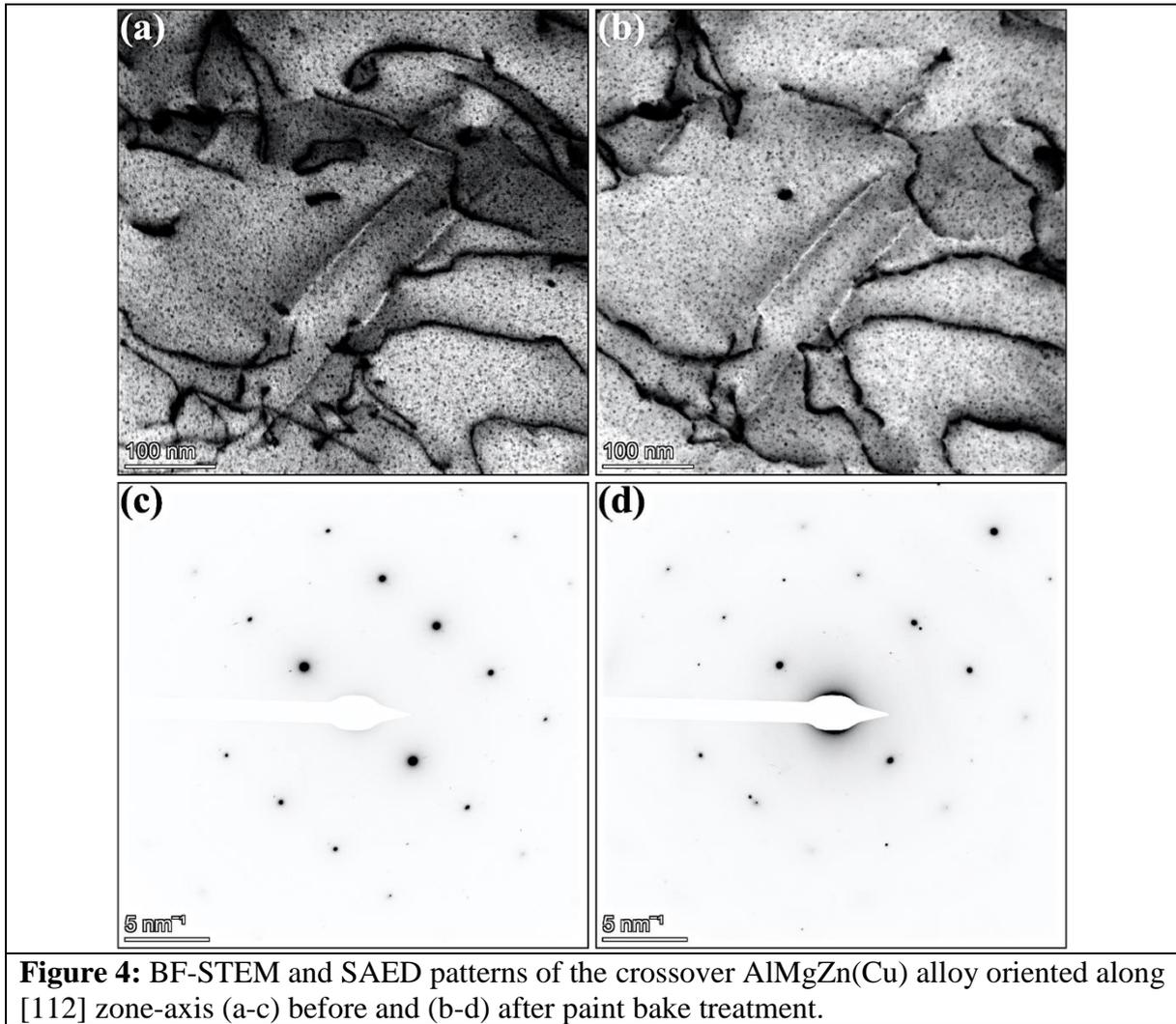

**Figure 4:** BF-STEM and SAED patterns of the crossover AlMgZn(Cu) alloy oriented along [112] zone-axis (a-c) before and (b-d) after paint bake treatment.

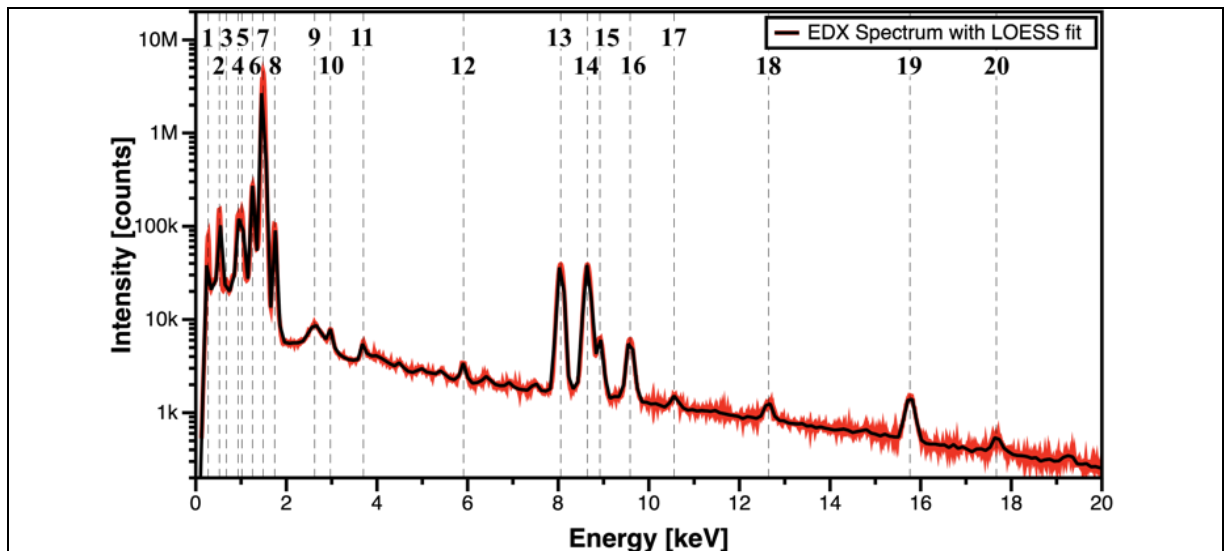

**Figure 5:** STEM-EDX raw spectrum acquired from the whole area covered in micrograph 4(b). Note: the y-scale was set to logarithm in order to maximize visualisation smaller peaks.